\documentclass[12pt,preprint]{aastex}
\newcommand{\brr}{$\langle B_{||} \rangle_{r=20}$}
\newcommand{\soho}{{\em SOHO{}}}
\newcommand{\pref}{\protect\ref}

\begin{document}

\shorttitle{Loading EUV BrightPoints}
\shortauthors{S.~W. McIntosh}
\title{On The Mass and Energy Loading of EUV Brightpoints}
\author{Scott W. McIntosh}
\affil{Department of Space Studies, Southwest Research Institute,\\
1050 Walnut St, Suite 300, Boulder, CO 80302}
\affil{High Altitude Observatory, National Center for Atmospheric Research,\\
P.O. Box 3000, Boulder, CO 80307}
\email{mcintosh@boulder.swri.edu}

\begin{abstract}
We discuss the appearance of EUV brightpoints (BPs) in the analysis of long-duration observations in the \ion{He}{2}~304\AA{} passband of the Solar and Heliospheric Observatory (SOHO) Extreme-ultraviolet Imaging Telescope (EIT). The signature of the observed 304\AA{} passband intensity fluctuations around the BPs suggest that the primary source of the mass and energy supplied to the magnetic structure is facilitated by relentless magnetoconvection-driven reconnection, forced by the magnetic evolution of the surrounding supergranules. Further, we observe that if the magnetic conditions in the supergranules surrounding the footpoints of the cool 304\AA{} BPs are sufficient (large net imbalance with a magnetic field that closes beyond the boundaries of the cell it originates in) the magnetic topology comprising the BP will begin to reconnect with the overlying corona, increasing its visibility to hotter EUV passbands and possibly Soft X-Rays.
\end{abstract}

\keywords{Sun: granulation \--  Sun:magnetic fields \-- Sun: chromosphere \-- Sun:transition region \-- Sun:corona}

\section{Introduction}
Recent results \citep[][]{McIntosh2006a, McIntosh2006b, Jefferies2006} have provided detailed observational support for the widely held hypothesis that the relentless action of magnetoconvection-driven reconnection \citep[e.g.,][]{Priest2002} supplies the bulk of the energy to the solar chromosphere and transition region through thin ejecta that are intrinsically tied to supergranular spatial scales. These observational results were derived from the multi-thermal analysis of observations from the Solar and Heliospheric Observatory \citep[\soho,][]{Fleck1995} and timeseries analysis of data from the Magneto-Optical Filter at Two Heights \citep[MOTH;][]{Finsterle2004a}, and suggest that the small-scale eruptive phenomena observed are linked to the ubiquitous appearance of ``spicules'' at the solar limb in eclipse or coronagraph observations \citep[e.g.,][]{Secchi1877, Roberts1945}. These parallel analyses indicate that the spicular activity is originated by forced magnetic reconnection which gives rise to the evaporative mass loading and initial energy deposition in the newly created topology. The creation of this new magnetic topology locally modifies the acoustic cut-off frequency \citep[][]{McIntosh2006} and permits {\it p-mode} leakage, propagation, energy dissipation \citep[]{Jefferies2006} and mass transport along the inclined field structure \citep[see e.g.,][]{DePontieu2004a, Hansteen2006, DePontieu2007,DePontieu2007b}. In short, the data are consistent with an energy and mass release process that combines driven magnetic reconnection and magneto-acoustic wave leakage and dissipation in one elementary structural unit (at a spatial resolution of 5\arcsec{}). It is possible that this unit is a jet-like ``spicule'', and these spicules appear to be systematically distributed over the solar disk in and around the granular and supergranular network. Throughout we will use the term spicule to describe these impulsive jet-like phenomena observed with the provision that we do not have the coverage of other observations (simultaneous observations in H$\alpha$ or \ion{Ca}{2}) to unambiguously link these jets with classical spicules, fibrils or mottles. The coupling of the extreme-ultraviolet and optical phenomena is clearly complicated, is a topic beyond the scope of this paper, but one that will be explored in a forthcoming paper (McIntosh, Judge \& Gurman 2007 \-- in preparation).

The analysis presented here \citep[like that of][\-- hereafter Paper 1]{McIntosh2007}, was motivated by a movie presented on the \soho{} internet site\footnote{The movie can be viewed at {\url http://sohowww.nascom.nasa.gov/pickoftheweek/old/24mar2006/}.} from the Extreme-ultraviolet Imaging Telescope \citep[EIT;][]{Boudine1995}. This movie shows the Sun ``percolating'' with intensity brightenings in the 304\AA{} (\ion{He}{2}) passband that is formed at the interface between the chromosphere and transition region \citep[][Paper~1]{JudgePietarila2004, PietarilaJudge2004}. These brightenings occur on spatial and temporal scales that are commensurate with the phenomena discussed in the preceding paragraph, strongly suggesting that they are the result of the same physical mechanism. To that end, an analysis of these 304\AA{} intensity brightenings provides a means of testing the magnetoconvective energy and mass loading hypothesis that is presented in \citet{McIntosh2006b}. This Paper follows directly from the analysis of Paper 1 and provides evidence supporting the dominance of high-$\beta$ forced mass and energy loading in the creation and evolution of X-Ray, EUV Bright Points \citep[BPs; see e.g.,][]{Golub1974, Habbal1981, Harvey1997} and \ion{He}{1} 10830\AA{} ``Dark Points'' \citep[e.g.,][]{Harvey1994a,McQueen2000}.

In the following section, we will briefly discuss the time-series observations and basic analysis of the 304\AA{} (\ion{He}{2}) passband lightcurves that form the basis of our study. In Sect.~\pref{results} we will investigate the spatial and statistical dependence of the \ion{He}{2} 304\AA{} intensity light-curves in and around a set of sample BPs and a setof BPs identified by an automated algorithm \citep[][]{McIntoshGurman2005} designed for that purpose. In Sect.~\pref{discuss} we discuss the results of the analysis, placing them in context with previous investigations relating BPs and spicules \citep[e.g.,][]{Moore1977, Rovira1999, Madjarska2002} and their connection to the coronal \citep[e.g.,][]{Longcope2001, Brown2001, Brown2002} and photospheric magnetic field \citep[][]{McIntosh2006a}. 

We will see that the mass and energy ``loading'' of the cool BPs (those visible only in the 304\AA{} timeseries) comes from the interaction of the relentlessly emerging and advecting magnetic flux in the vicinity of magnetic dipoles that form its structural core and can therefore explain the appearance of dynamic impulsive heating \citep[e.g.,][]{Habbal1981, Porter1987, Habbal1988, Habbal1991, Rovira1999, McQueen2000, Urra2004b} and simultaneous observation of running low-frequency ($\le$5mHz) waves \citep[e.g.,][]{Urra2004,Urra2004a} in the BP structure. Further, the results of our analysis are consistent with a two-stage BP loading mechanism that can go some way to providing the ``necessary and sufficient'' conditions \citep[][]{Harvey1994b} for the production of a hot BP (those observed in hotter EUV passbands and soft X-Rays). We infer that forced magnetoconvection-driven reconnection occurs in all cool BPs, but that increased energy supply from the magnetic environment surrounding the BP causes an eventual interaction of the BP magnetic flux with overlying coronal magnetic flux, producing hot BPs.

\section{Observations and Analysis}\label{secobs}
The results presented in this Paper are based on the analysis of a long-duration sequence of EIT 304\AA{} passband imaging, starting 2006/03/18 20:36UT and ending 2006/03/21 23:48UT. The subset of the EIT sequence analyzed here has a fixed cadence of twelve minutes, no missing data blocks (2006/03/19 00:11 \-- 2006/03/19 23:59), and covers the approximate evolutionary lifetime of a supergranular cell \citep[$\sim$20 hours][]{Harvey1994a}. The selected subfield of the EIT data sample used in this analysis (reduced using the standard package discussed in the \anchor{http://umbra.nascom.nasa.gov/eit/eit_guide/}{EIT User's Guide}) has been tracked, extracted and co-aligned using the Interactive Data Language solar data mapping software in the SolarSoft data analysis tree \citep{Freeland1998}. We estimate \citep[using a cross-correlation algorithm used to obtain sub-pixel timeseries co-alignment with TRACE data, e.g.,][]{McIntosh2004} that the data are tracked and co-aligned to within half of a pixel over the period needed, using the middle image (taken at 13:19UT) as the reference.

The 304\AA{} sequence is augmented by coronal context images from the EIT 195\AA{} passband and full disk line-of-sight magnetograms ($B_{||}$) from the Michelson Doppler Imager \citep[][]{Scherrer1995} taken closest to the start of the sequence, 01:12UT and 01:36UT respectively. In later sections we will also use the EIT 195\AA{}, 171\AA{} and 284\AA{} synoptic images taken at six-hour intervals over the course of the day (01:00, 07:00, 13:00, 19:00UT) to identify visible coronal BPs \citep[][]{McIntoshGurman2005}.

\begin{figure}
\epsscale{0.9}
\plotone{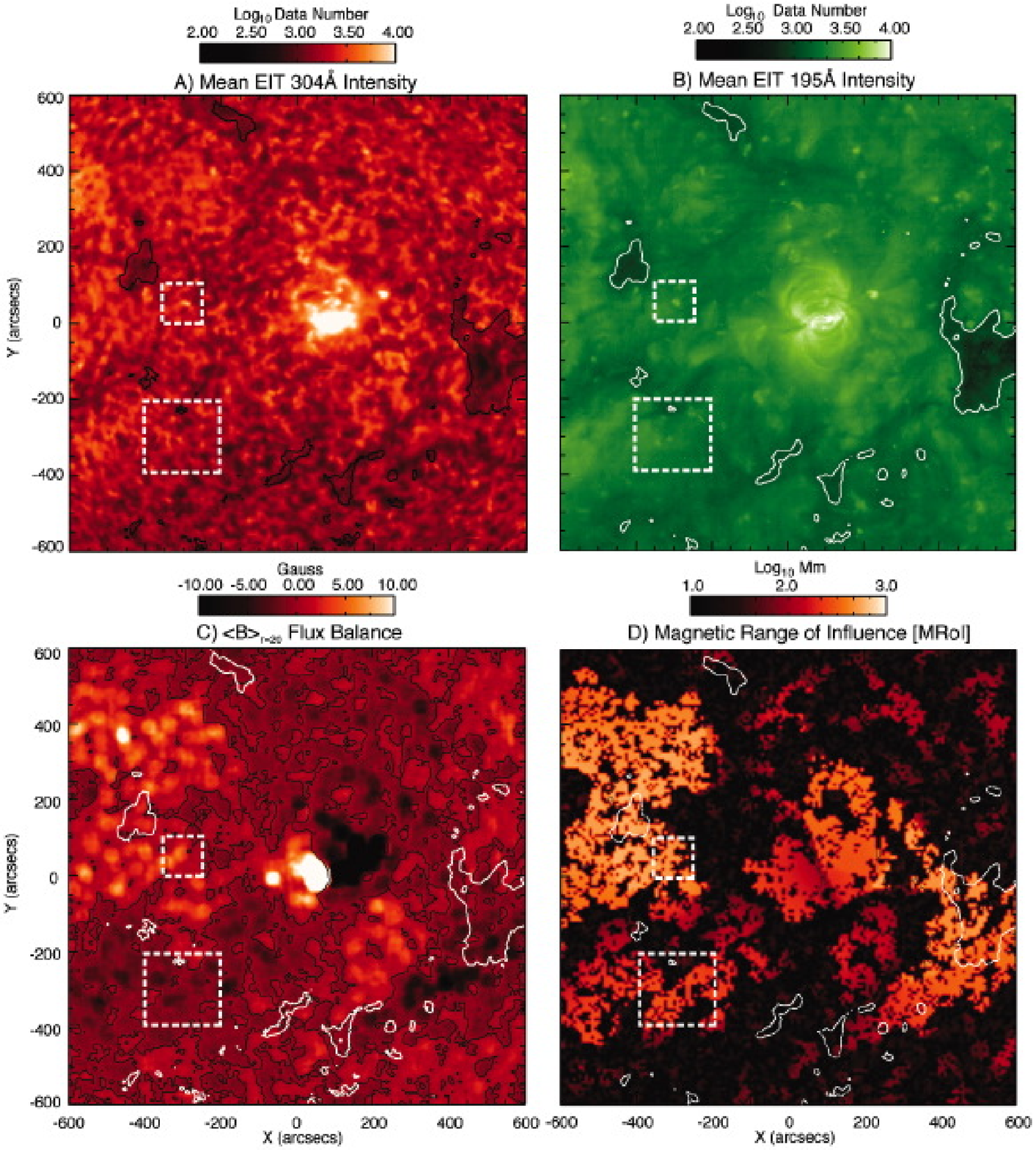}
\caption{Context maps for the 2006 March sequence. We show the mean EIT 304\AA{} intensity (panel A), 195\AA{} coronal context image (panel B), the supergranule averaged magnetic field (panel C) and Magnetic Range of Influence (panel D) where the latter (\brr and MRoI) were developed in \citet{McIntosh2006a}. In each panel we show the EIT 195\AA{} 150 Data Number intensity contour (from panel B) as a proxy of coronal hole boundaries \citep{McIntosh2006a} and the dashed square regions that isolate the sample BPs detailed in Figs.~\pref{bp1} and~\pref{bp2} respectively. The black contour in panel C shows the location of the large scale magnetic neutral line \brr=0~G. \label{contextmarch2006}}
\end{figure}

Figure~1 of Paper~1 shows the full-disk context images and subfield pointing for the sequence that we examine here. That figure shows the EIT 195\AA{} coronal image (panel A) , the first EIT 304\AA{} image in the sequence (B), the MDI $B_{||}$ image (C) and, in panel D, a smoothed $B_{||}$ image \citep[\brr;][]{McIntosh2006a} providing information about the net-magnetic flux balance on the supergranular scale (i.e., the magnetogram is convolved with circular kernel of diameter 20Mm). The same figure also shows the 1200\arcsec x 1200\arcsec{} field of view that is used in the primary study of this paper (black square in panels A and B, white in panels C and D) and shown here in Fig.~\pref{contextmarch2006}.

In the following interpretation of this sequence we assume that the 304\AA{} passband consists primarily of emission from \ion{He}{2} formed at a temperature of roughly 50,000~K \citep[][]{Mazzotta1998}. Aside from the strong emission from the \ion{He}{2} 303.78\AA{} line, there is a contribution to the bandpass emission in an active region from the \ion{Si}{11} 303.31\AA{} line and the blended emission of \ion{Mn}{14} and \ion{Fe}{15} at 304.86\AA{} (all three contaminant lines are formed in the hotter corona at $\sim$1.5MK or above) in the ratio 10::1::0.06 \citep[][]{Brosius1998}. The measured contribution of \ion{He}{2} emission relative to the emission lines of \ion{Fe}{10}/\ion{Fe}{11} (i.e., the contribution from the EIT 171\AA{} passband in second spectral order at about $\sim$1MK) is 25 to 1 in an active region and 10 to 1 in the quiet sun \citep[][]{Brosius1998}. Further, the analyses of \citet{JudgePietarila2004} and \citet{PietarilaJudge2004} have demonstrated that the formation process for the \ion{He}{2} plasma observed in emission (and \ion{He}{1} 10830\AA{} in absorption) exhibits dynamic behavior that is characteristic of the upper chromosphere and the low solar transition region. This thermodynamic placement of \ion{He}{2} makes it a useful proxy of plasma dynamics in and around the ``magnetic transition'' region where the gas pressure and magnetic pressure of the plasma are in approximate balance; a nexus between the low and high\--$\beta$ plasma regimes.

\subsection{Data Analysis}\label{data}
The analysis of the EIT 304\AA{} data presented is very straightforward, is easily repeatable and has two parallel threads to explore the temporal and spatial behavior of the observations. In the first thread, we analyze the intensity light curves at each spatial pixel in the de-rotated, co-aligned and trend (a ten timestep box-car smoothed profile of the intensity) removed ($\hat{I}(x,y,t) = {I}(x,y,t) - <{I}(x,y,t)>_{10}$) datacube to count the number of significant brightenings over the observing period. Each brightening counts as one instance of an event and we define a significant brightening as a 10\% change in the trend-removed ratio $\Delta \hat{I} /  \hat{I}$ (= [$\hat{I}(x,y,t-1) - \hat{I}(x,y,t)] / \hat{I}(x,y,t)$), where $t$ is measured in units of the (12 minute) time-step between frames of the observing sequence. The second thread investigates the spatial dependence of the brightenings through the pixel-by-pixel intensity distribution, $f(I(x \pm \delta x, y \pm \delta y,t))$, where $\delta x$ and $\delta y$ are each two pixels. Using an extended spatial range in the analysis ensures that the intensity distribution has a sufficient number of elements to ensure that the moments of the distribution and fits to its shape can be accurately computed. This way of accumulating the data explicitly assumes that the plasma in the adjoining pixels is physically coupled to that in the pixel being studied. Section~3 of Paper~1 defines and discusses the pixel light curve distribution diagnostics used in the present analysis: the normalized distribution width (the width of the intensity distribution divided by the mean intensity in the pixel); distribution skewness; distribution kurtosis; and event ``number''. 

The event number diagnostic is incremented every time there is a 10\% change in $\Delta \hat{I} /  \hat{I}$ of the time series in the pixel while the skewness \citep[see, e.g.,][]{Abramowitz1972,Press1992} of the intensity distribution measure whether outliers in the intensity distribution are below (negative) or above (postive) the distribution mean, skewing it to the left or the right respectively. The kurtosis of the distribution indicates whether the shape of the distribution is more or less sharply peaked than a Gaussian (a kurtosis of zero is perfectly Gaussian) while a flat value has a negative kurtosis and very sharply peaked distributions result in a positive kurtosis.

\section{Results}\label{results}
\begin{figure}
\epsscale{0.9}
\plotone{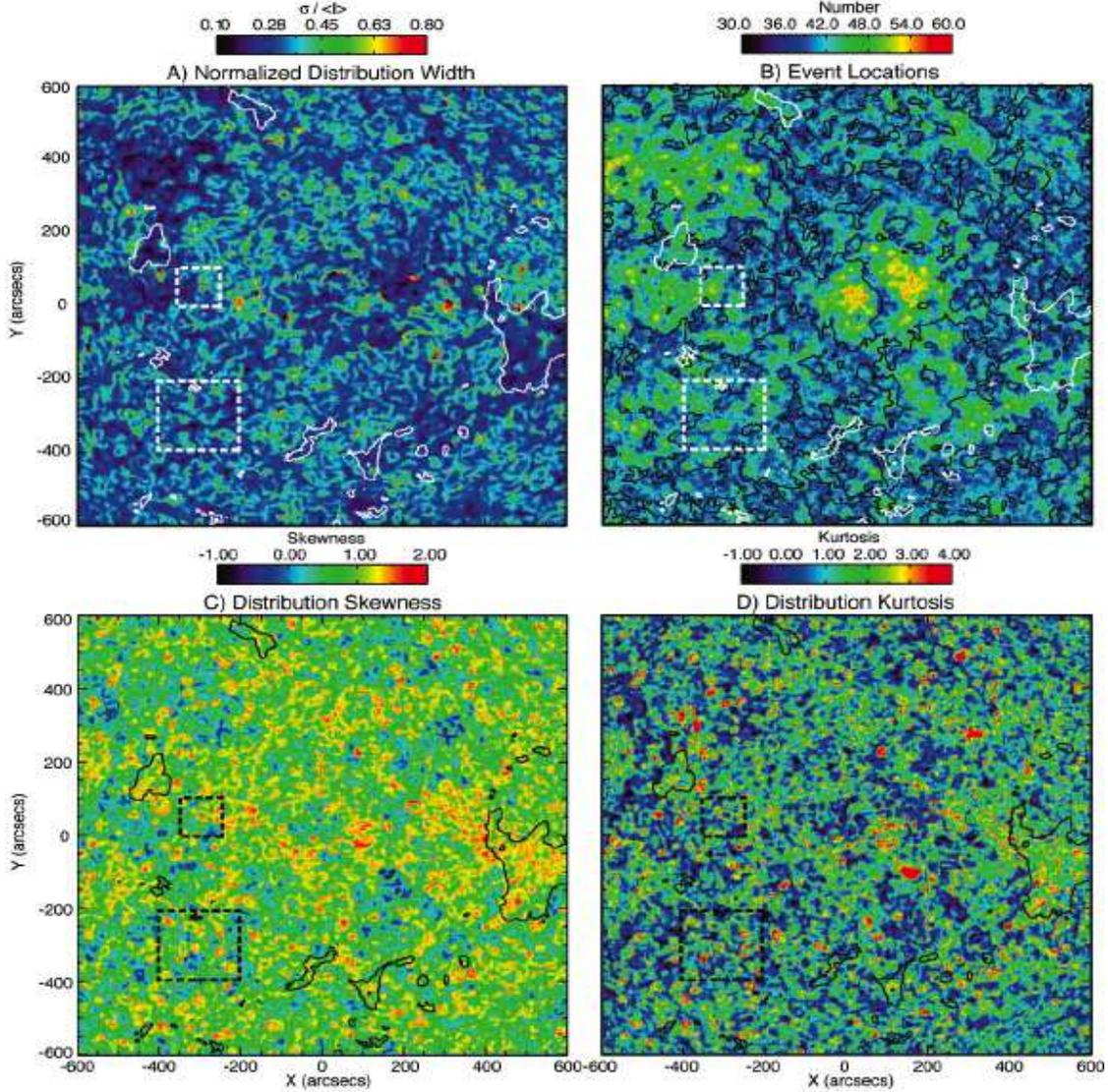}
\caption{Distribution diagnostic maps for the 2006 March sequence. We show the normalized intensity distribution width (panel A), intensity ``event'' count (panel B), distribution skewness (panel C) and kurtosis (panel D) maps. In each panel we show the EIT 195\AA{} 150 Data Number intensity contour (from panel B of Fig.~\pref{contextmarch2006}) as a proxy of coronal hole boundaries \citep{McIntosh2006a} and the dashed square regions isolate the sample BPs detailed in Figs.~\pref{bp1} and~\pref{bp2} respectively. The black contour in panels B shows the location of the large scale magnetic neutral line \brr=0~G, derived from panel C of Fig.~\pref{contextmarch2006}. \label{resultsmarch2006}}
\end{figure}

\begin{figure}
\epsscale{1.0}
\plotone{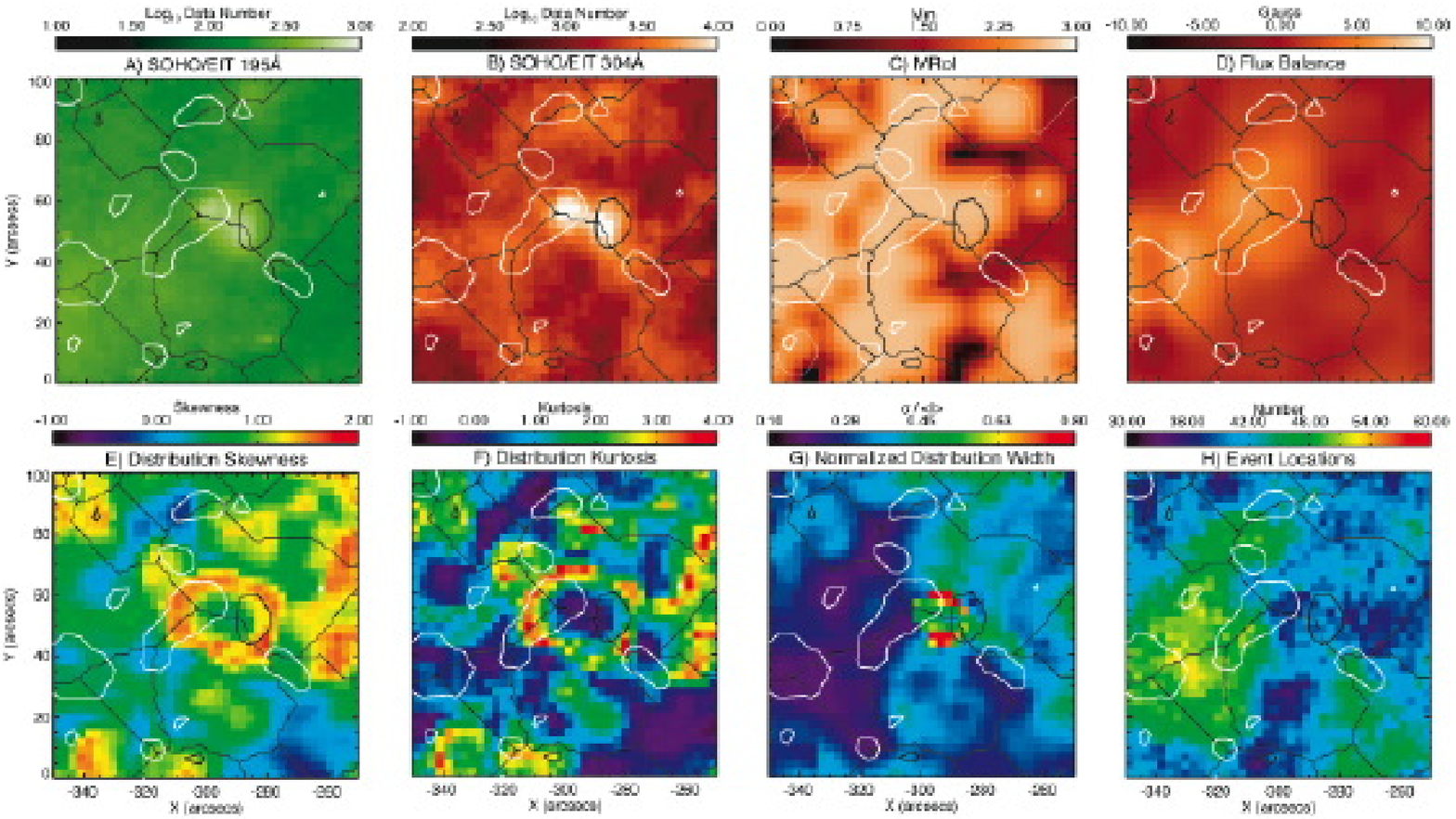}
\caption{The 100\arcsec{}x100\arcsec{} subfield of the March 2006 sequence to illustrate the typical statistical signature of a single EUV  brightpoint. The top row of panels show the EIT 195\AA{} context image (A), the 304\AA{} mean intensity (B), MRoI (C) and \brr{} (D) for the region while the bottom row shows the four intensity distribution diagnostic maps; skewness (E), kurtosis (F), normalized distribution width (G) and event number map (H). Each panel shows the watershed segmentation supergranular boundaries determined from panel B and the $\pm$ 25~G magnetic flux contours from the full-resolution SOHO/MDI magnetogram (also used to construct the MRoI and \brr{} maps). \label{bp1}}
\end{figure}

\begin{figure}
\epsscale{1.0}
\plotone{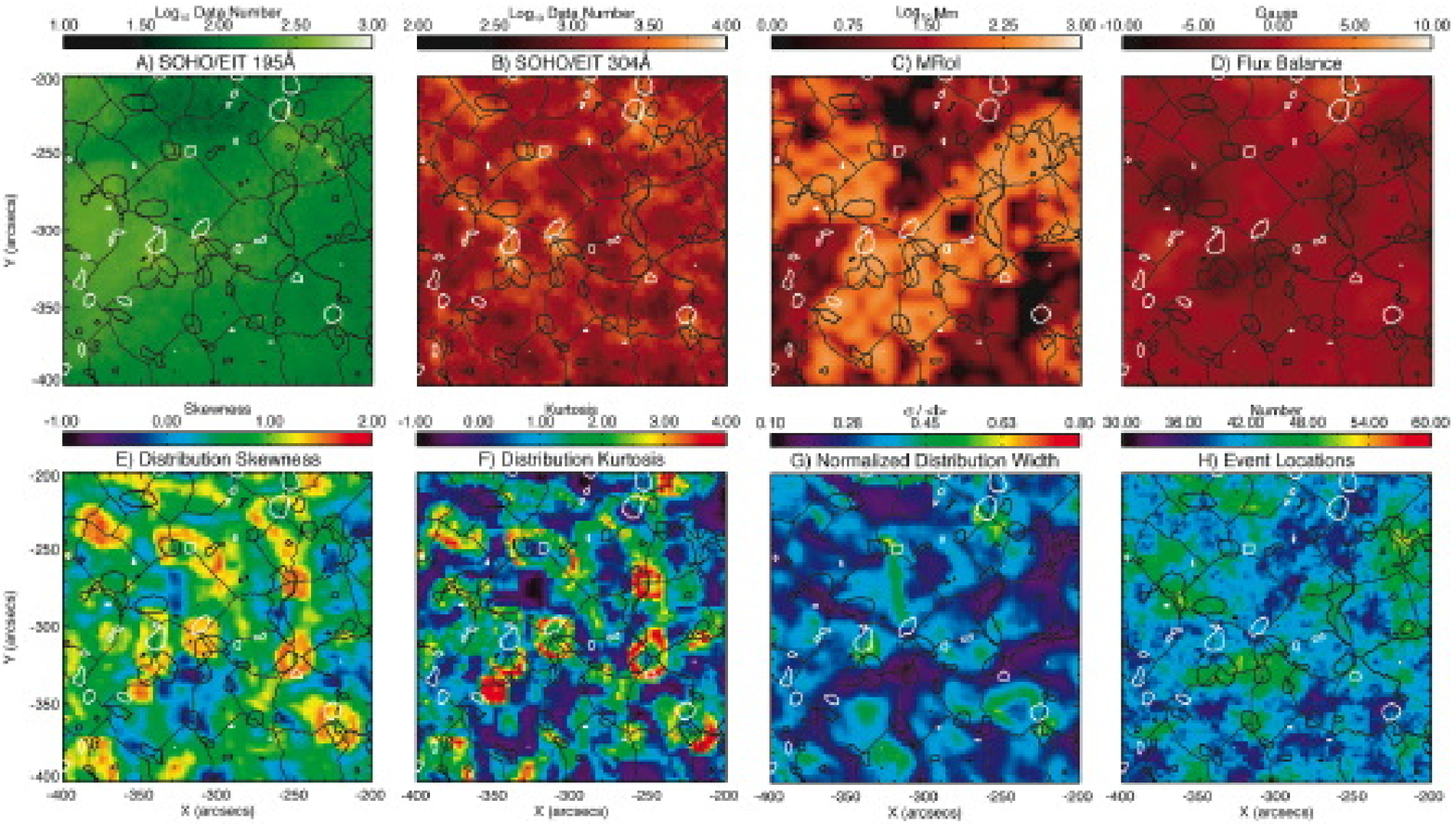}
\caption{The 200\arcsec{}x200\arcsec{} subfield of the March 2006 sequence to illustrate a ``cluster'' of EUV brightpoints. The top row of panels show the EIT 195\AA{} context image (A), the 304\AA{} mean intensity (B), MRoI (C) and \brr{} (D) for the region while the bottom row shows the four intensity distribution diagnostic maps; skewness (E), kurtosis (F), normalized distribution width (G) and event number map (H). Each panel shows the watershed segmentation supergranular boundaries determined from panel B and the $\pm$ 25~G magnetic flux contours from the full-resolution SOHO/MDI magnetogram (also used to construct the MRoI and \brr{} maps). \label{bp2}}
\end{figure}

Figure~\pref{contextmarch2006} shows the mean EIT 304\AA{} intensity over the observing sequence (panel A), the EIT 195\AA{} coronal image (panel B), the smoothed $B_{||}$ image (\brr, providing information about the net-magnetic flux balance on the supergranular scale \-- panel C) and the ``Magnetic Range of Influence'' (MRoI, indicating the distance from a given pixel needed to balance the flux contained there \-- panel D); introduced and discussed at length in \citet{McIntosh2006a}. Figure~\pref{resultsmarch2006} shows the intensity distribution diagnostics in the same 1200\arcsec{}x1200\arcsec{} region: the mean normalized distribution width (panel A), intensity ``event'' count (panel B), skewness (panel C) and kurtosis (panel D) maps. With close inspection of panels C and D the reader will see a prevalent pattern of circular features, or ``rings'', that are visible (at the same location) in the distribution skewness and kurtosis maps. These features are visible in the dashed square regions that are provided as Figures~\pref{bp1} and~\pref{bp2} and are discussed below. Much of the detail present in this figure is not discussed here, but is left for discussion in a future paper (McIntosh, Judge \& Gurman 2007 \-- in preparation).

Panels E and F of Fig.~\pref{bp1} demonstrate the presence of a very distinct ring signature in the skewness and kurtosis diagnostic maps. In concert with the kurtosis and skewness rings present in panels~E and~F we see that a significant enhancement in the normalized distribution widths (panel G) that sit on the line geometrically bisecting the two magnetic flux concentrations that mark the footpoints of the BP (show as black \--negative \-- and white \--positive\-- contours from the full resolution SOHO/MDI magnetogram at a level of 25G respectively). We can see that the BP structure (in 195\AA{} and 304\AA{}; panels A and B) is outlined by these rings. The prevalent connection of the skewness and kurtosis ring signature (cf. Fig.~\pref{resultsmarch2006}) may provide some insight into how BPs are loaded with mass and heated. Recall, from Paper~1, that we have interpreted increases in distribution skewness and kurtosis as being due to small, but impulsive brightenings in $\delta$I/I that are significantly greater than the mean intensity level in the pixel. However, these brightenings may not always be of large enough magnitude in and around these bright structures to constitute the detection of an event (panel H shows a very weak enhancement in the number of events surrounding the magnetic dipole in the ring location). 

We know that some portion of the emerging magnetic flux in the interior of the supergranular cells, whose vertices comprise the BP's magnetic footpoints, is advected towards the boundary connecting them. It is likely that the advecting flux elements are forced to reconnect with the magnetic topology in whatever way they can, impulsively changing the intensity surrounding the BP and creating the skewness and kurtosis rings while they load mass and thermal energy into that topology \cite[cf. the quiet Sun heating and mass loading proposed by][]{Jefferies2006, McIntosh2006b, DePontieu2007b}. In addition, the global flow-field will continue to push and pull the BP footpoints at random giving an additional magnetic energy impulse to the complex system as the supergranular cells are relentlessly being created and destroyed. The BP will continue to consume the emerging/advected flux that is heading in its direction, heating and loading until is stops emitting and dies. In the much the same way that the BP is born, its death is most likely to occur with the footpoint magnetic flux eventually cancel with one another (see Sect.~\pref{single}). Because the BP footpoints are located at vertices of the same, or neighboring, supergranular cells they are separated by, at most, a supergranular diameter ($\sim$20Mm). Therefore, the lifetime of the BP will depend on the evolutionary lifetime of the separating cells. If they collapse, the BP footpoints will coalesce and eventually cancel each other's magnetic flux. Conversely, if the intermediate supergranules do not collapse, then the BP will continue to exist and be loaded by the advecting flux surrounding it. A detailed understanding of BP lifetimes can be undertaken with a much larger dataset \citep[][]{McIntoshGurman2005}, but is beyond the scope of this paper and left for future work (McIntosh \& Davey 2007 in preparation).

From Fig.~\pref{bp2} we make an important distinction about the physical connection between BPs and skewness/kurtosis rings. While each skewness/kurtosis ring has a bright 304\AA{} emission and a magnetic foundation (at least a dipolar structure) at it's core not every skewness/kurtosis ring has a ``hot'' EUV BP associated with it (i.e., bright emission in a ``true'' coronal passband, e.g., EIT 171, 195, 284\AA{} or Soft X-Rays). This fact has already been noted in the BP literature \citep[e.g.,][]{Porter1987,Harvey1993,Harvey1994a,Harvey1994b} where it was clearly demonstrated that not all ``ephemeral active regions'' have hot BPs associated with them. Comparing the panels of Fig.~\pref{bp2}, we see that the brightest emitting regions in panel A (that overlie magnetic dipoles) have associated bright 304\AA{} emission in panel B. Further, these locations also have large MRoIs and (at least) a moderate \brr{}. Of the seven locations with skewness and kurtosis rings [centered on (-330, -250), (-340,-310), (-310,-310), (-250,-320), (-250,-220), (-220, -360), (-250, -240)] five have bright coronal emission associated with them [(-330, -250), (-340,-310), (-310,-310), (-250, -220), (-250, -240)]. 

There are two regions where no coronal BPs exist even though the dipoles are in place and there is (reasonably) bright emission in the 304\AA{} passband ((-220, -360) and (-250, -320)); these regions are examples of small BP ``cavities''. The rings in these regions occur where the MRoI is small in one or both footpoints and the \brr{} is low in the vicinity of the BP. We believe that the presence of cavities tells us a great deal about the mass and energy loading of the individual dipoles and that it must be related to the magnetic conditions in {\em and} around the dipole \citep[see above and][]{Harvey1994a} \-- not all of these rings/dipoles receive the same mass and energy flux and so we observe the fact that not all (EIT detected) BPs occur in all coronal passbands, or different temperature images see different numbers of BPs \citet[][]{McIntoshGurman2005}. In Sect.~\pref{compare} we will compare the locations of skewness/kurtosis rings in Fig.~\pref{resultsmarch2006} with the BP locations automatically detected in the three other EIT passbands to study the properties needed to create a BP cavity.

\subsection{The Evolution of One 304\AA{} Brightpoint}\label{single}
\begin{figure}
\plotone{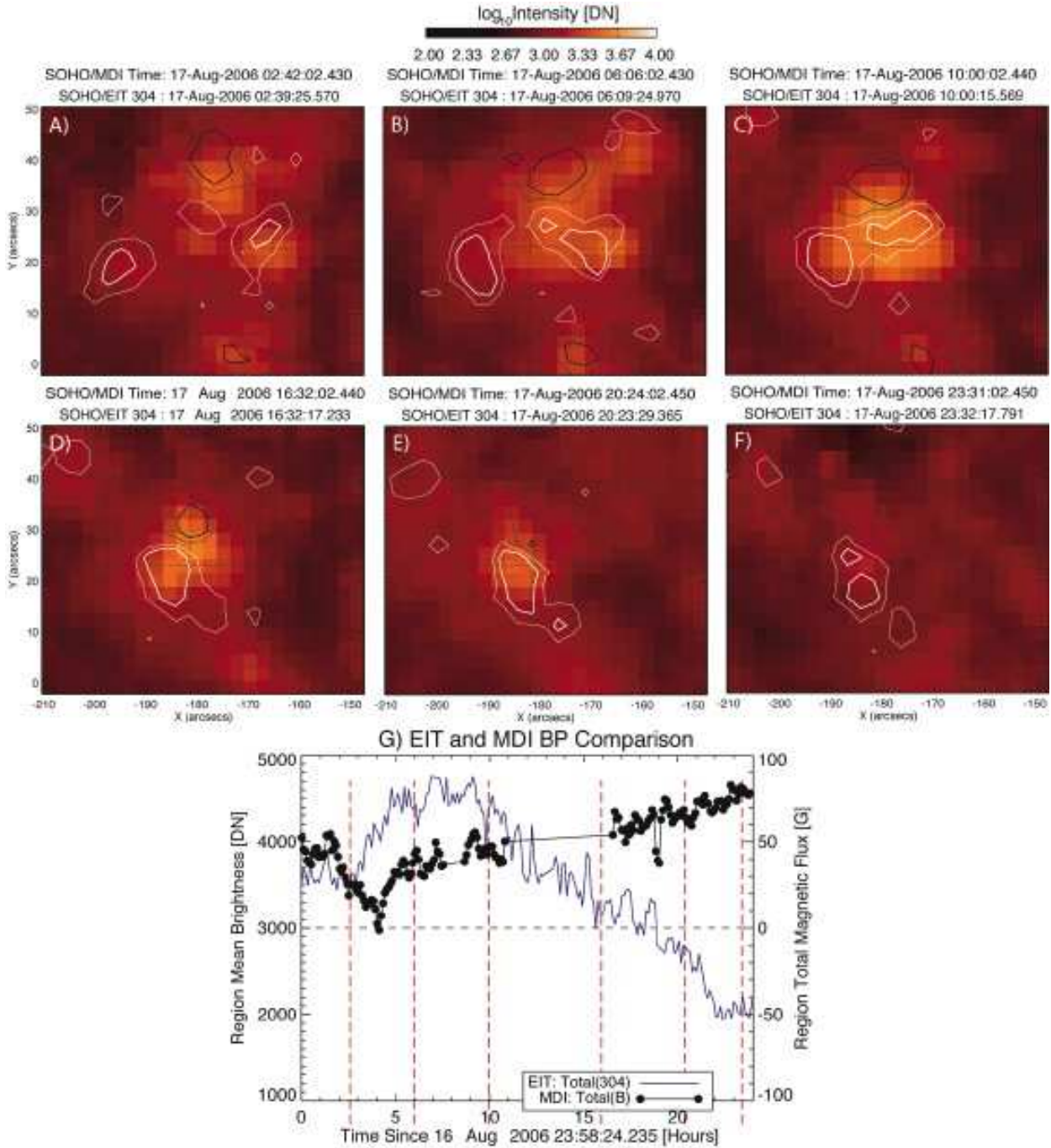}
\caption{The evolution of a single 304\AA{} BP from the August 2006 sequence. The perimeter panels show snapshots of the emitting BP structure and the SOHO/MDI magnetograms nearest to those frames. The thin and thick contours represent field strengths of 10 and 25~G respectively. Panel G compares the totaled 304\AA{} intensity (blue lightcurve) and the total magnetic flux (black dots) in the region \-- the vertical red dashed lines mark the times of the perimeter panels. The online edition of the journal has a movie based on panels A through F of this figure. \label{res4}}
\end{figure}

We have seen in the preceding figures and paragraphs that the EUV BP has a very particular statistical signature (as far as our analysis is concerned). Figure~\pref{res4} provides evolutionary snapshots, from birth to death, of one single 304\AA{} BP extracted from three day EIT sequence at 6 minute cadence, starting 23:58UT on 2006, August 16 and ending at 23:43UT on 2006, August 18 \citep[this sequence will be studied in some detail by][]{McIntosh+others2007b}. This EIT sequence is very useful for looking at the detailed evolution of the BP and also has the benefit of 1-minute cadence full-disk MDI magnetogram observations, for the large part August 17. With these data we can monitor the evolution of the BP over its lifetime with respect to the magnetic field creating and feeding it. In the panels A-F of Figure~\pref{res4} we show the instantaneous EIT 304\AA{} intensity and the nearest MDI magnetogram to that time (the MDI data shown are the result of a six-minute average of the full-disk one-minute magnetograms), the thick black and white contours represent a magnetogram levels of -25 and 25G while the thin contours are $\pm$10G, respectively. For comparison with the frames presented in panels A through F we show the evolution of the mean brightness lightcurve and the total magnetic flux of the available MDI magnetograms over the whole 60\arcsec{}x50\arcsec{} region surrounding the BP in panel g, shown as a solid blue line and black joined dots respectively. The online version of this Paper has a movie illustrating the lifetime of the BP shown.

In panel A there are four main flux elements in the region around the BP that is beginning to form. As the large negative flux concentration at the top of the panel migrates Southward and the large positive concentration on the left migrates Eastward the net magnetic flux in the region approaches zero ($\sim$04:00UT). The emission around the BP has already begun to increase significantly as we move into panel B and the migrating negative flux is canceling rapidly, evolving to a tripolar magnetic configuration and the BP is detected in EIT 195\AA{} image ($\sim$1.5MK plasma) at 07:13UT, reaching peak 304\AA{} brightness about then and staying there through panel C. Eventually, some six hours later, in panel D, the tripolar configuration has whittled down again to a N-S dipole configuration with the coalescing of the evolving positive flux elements and some significant removal of negative flux through cancellation. By 20:00UT (panel E) most of the negative flux has been cancelled and the region is taking on a distinctly unipolar look, by 23:00UT you would not know that a BP had ever existed in the region. 

Over the course of 24 hours we have monitored the interaction of the magnetic flux in the region and the supergranular flow it is anchored in, as it creates and destroys an EUV BP. This is a very typical evolutionary path for a BP, where lifetimes of $\sim$20 hours are commonplace \citep[e.g.,][]{Harvey1994a}, tying their lifespan directly to the magnetic aggregation process driven by the converging supergranular flow patterns \citep[e.g.,][]{Crouch2007}. 

\begin{figure}
\plotone{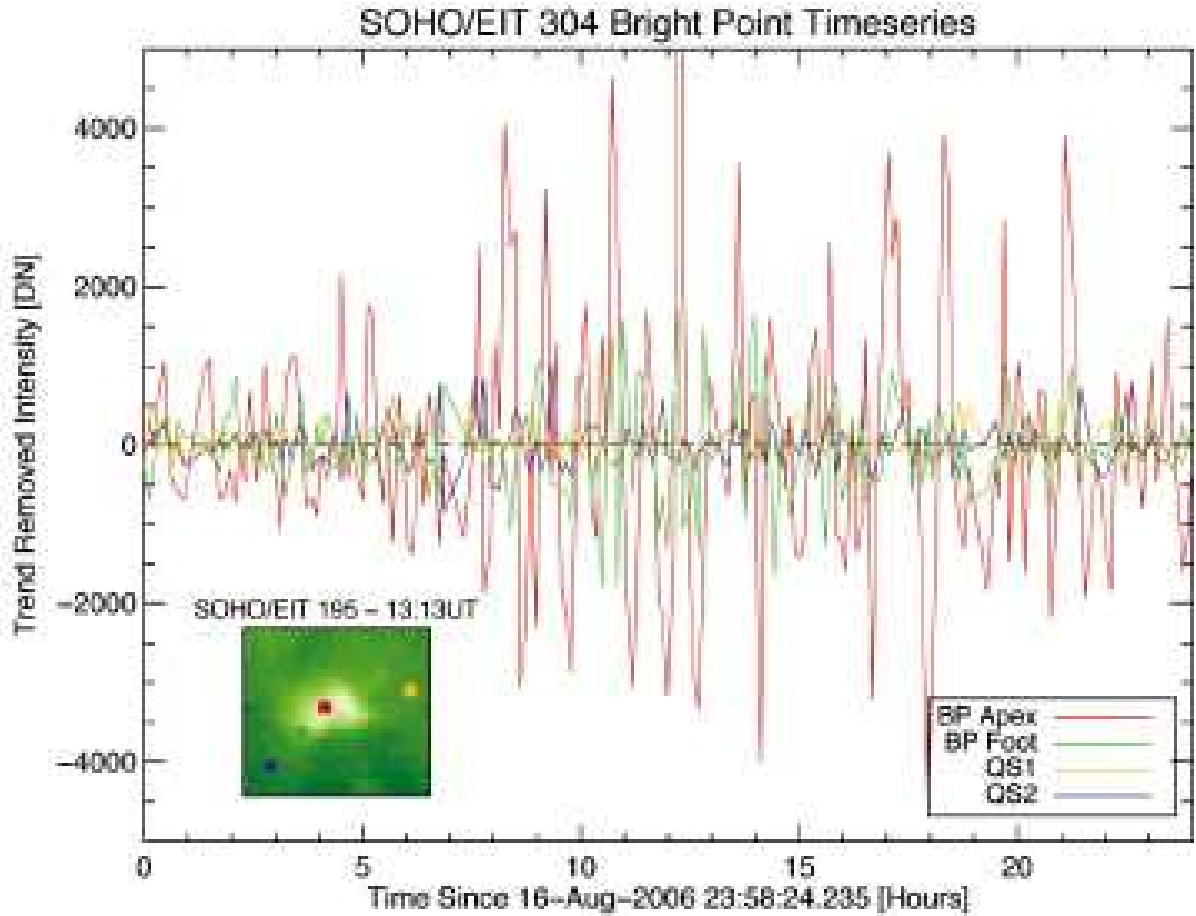}
\caption{Trend-removed lightcurves of individual pixels in the region of the BP shown in Fig.~\pref{res4}, see inset for reference colors. \label{res5}}
\end{figure}

\begin{figure}
\plotone{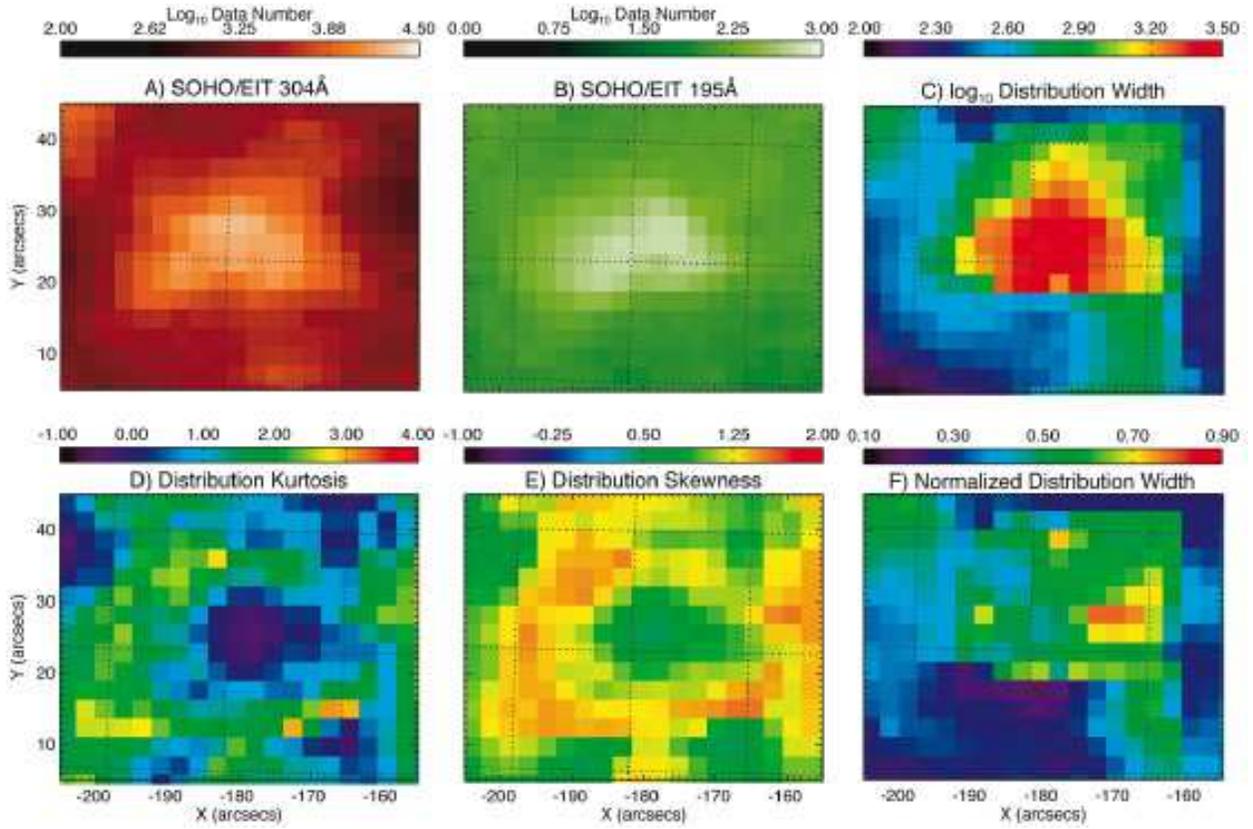}
\caption{Intensity distribution diagnostic maps for the BP region studied in Fig.~\pref{res4}, cf. Figs 2, 3 and 4. \label{res6}}
\end{figure}

In Figure~\pref{res5} we take a closer look at the BP and the fluctuations of the (trend-removed) 304\AA{} intensity at four points in the region (one in a footpoint \-- green; one near the apex of the structure \-- red; and two points in the surrounding plasma \-- blue and yellow \-- see inset image of the BP). The purpose of this figure is to demonstrate the range of variation in the 304\AA{} emission in and around the BP and to give us a feel for the influence these variations have on the distribution diagnostics that are the focus of this paper, and are shown for this BP, in Fig.~\pref{res6}. In the BP apex timeseries (red) we can see the time when the BP is reaching ``maturity'' (around 05:00UT) where the intensity fluctuations significantly increase in amplitude, but are not much higher than the mean intensity there ($\sim$4000DN) producing a broad distribution that has a kurtosis near zero and small positive skewness. In the quiet Sun surrounding the BP (blue and yellow) we see little evidence of the massive \ion{He}{2} brightenings that are visible in the BP apex (or footpoint) timeseries. However, there are a number of places in these timeseries where the change in intensity is just above the mean quiet Sun intensity ($\sim$300DN) and a probably significant enough in number to increase the skewness and kurtosis of the intensity distribution there. These intensity changes are, most likely, caused by the migration of the various flux elements that comprise the BPs structural core over its many phases of existence. The BP footpoint timeseries (green) is a hybrid of the evolution observed in the quiet Sun and BP apex cases, slightly higher enhancements in skewness and kurtosis (over the BP apex) and a relatively broad intensity distribution. Therefore, we see that the evolution of the magnetic flux (anchored in the high-$\beta$ plasma flow), that comprises the structure of the BP results, is consistent with the production of enhanced skewness and kurtosis in the intensity distributions that surround the brightpoint.

\subsection{Comparing Rings Locations with Automated EUV BP Detection}\label{compare}

We have seen above that not all BPs that are visible in 304\AA{} have a counterpart in the considerably hotter 171, 195 or 284\AA{} EIT passbands, i.e., there are BP ``cavities''. This is a correspondence first noticed by \citet[][]{Harvey1994b} who stated in their conclusion that ``the emergence or cancellation of magnetic fields in the photosphere is not in itself a necessary and sufficient condition for the occurrence of an XBP'', a X-Ray, or very hot BP. Fortunately, we have a tool to automatically detect BPs in the hotter EIT passbands \citep[][]{McIntoshGurman2005} that can be readily applied to all of the other EIT images taken on 2006 March 18. The detected BP locations can then be compared with the distribution and magnetic field diagnostic maps to see if a specific pattern appears; this pattern may shed some light on the necessary condition to produce an hot BP.

\begin{figure}
\epsscale{0.7}
\plotone{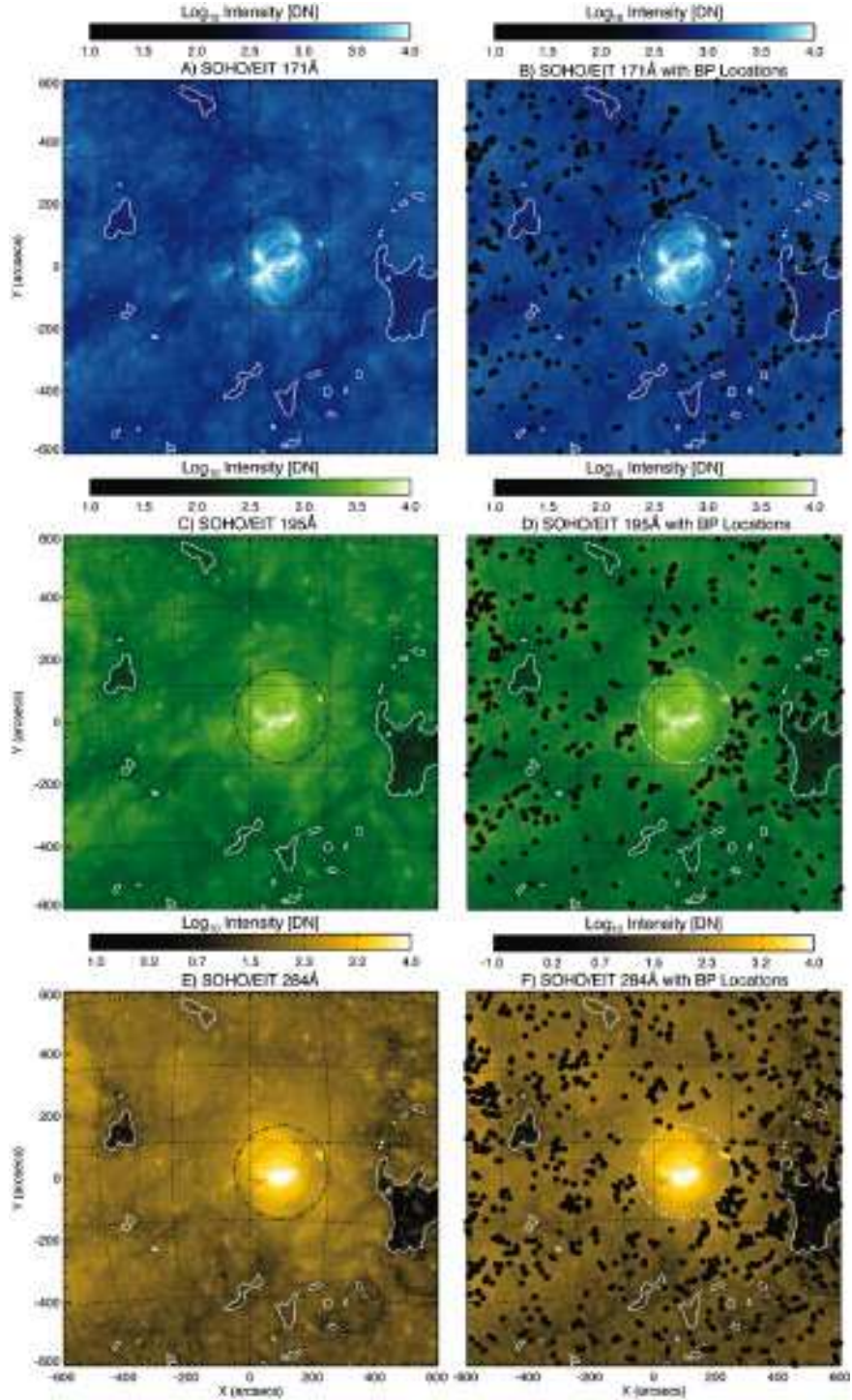}
\caption{Comparing the EIT SYNOP images (taken around 13:00UT 2006 March 19) with the detected EUV (hot) BPs (black dots) over the whole 24 hour period studied. From top to bottom we show the 171\AA{} with emission peaking at a temperature of about 1MK, 195\AA{} (1.5MK) and 284\AA{} (2.0MK). Each panel shows the coronal hole boundaries (white solid contour determined from the 195\AA{} image) and the circle dot-dashed surrounding the active region - we do not show the BPs detected in the circled region. \label{bp3}}
\end{figure}

On 2006 March 18 there were 4 images each in the 171, 195 and 284\AA{} EIT passbands taken around 01:00, 07:00, 13:00 and 19:00UT \-- the typical EIT ``SYNOP'' cadence. Since we have no information in the distribution diagnostics about when the skewness and kurtosis rings developed we must look at all of the BPs detected on that day to make our comparison. The left column of Fig.~\pref{bp3} (top to bottom) show the detected 171, 195 and 284\AA{} passband images taken around 13:00UT while the right column of panels show the corresponding BP detections for the whole day. Again, particularly in 171 and 195\AA{}, we see the cavities where very few BPs exist. The 284\AA{} passband has a contribution from transition region (\ion{S}{5}) network emission and may lead to a previously unnoticed large number of additional detections in that passband from a modification to the detection algorithm (McIntosh \& Davey 2007). All of the EIT images and BP locations are rotated to the mid-point of the 304\AA{} timeseries 13:19UT. We do not show the BP locations in the active region, just North-West of disc center (circled in the panels). Similarly, we acknowledge that the sparse 6-hour SYNOP observations may mean that we miss some hot BPs that occur, appearing and disappearing between images. This is not likely to be many as the mean BP lifetimes are $\sim$20 hours, but the amount missed can be estimated by comparing BP appearance rates and lifetimes at particular latitudes and longitudes, at different cadence of EIT images, using studies like that of \citet{McIntoshGurman2005}.

There are regions in the panels that show locations where there are very few, or no hot BPs, over the 24 hour period searched. These locations, like the lower left corner running diagonally upward towards disc center, are exactly what we mean by a BP cavity; BPs in 304\AA{}, but not in 171, 195 or 284\AA. All of the visible cavity regions appear to have another thing in common, they have very weak quiet Sun coronal emission. The apparently weak emission has a bearing on the reduced number of BPs detected if the analysis of \citet{McIntosh2006a} is correct as it reflects the energy (and mass) input for quiet solar corona and its dependence on the background magnetic footprint (\brr).

\begin{figure}
\plotone{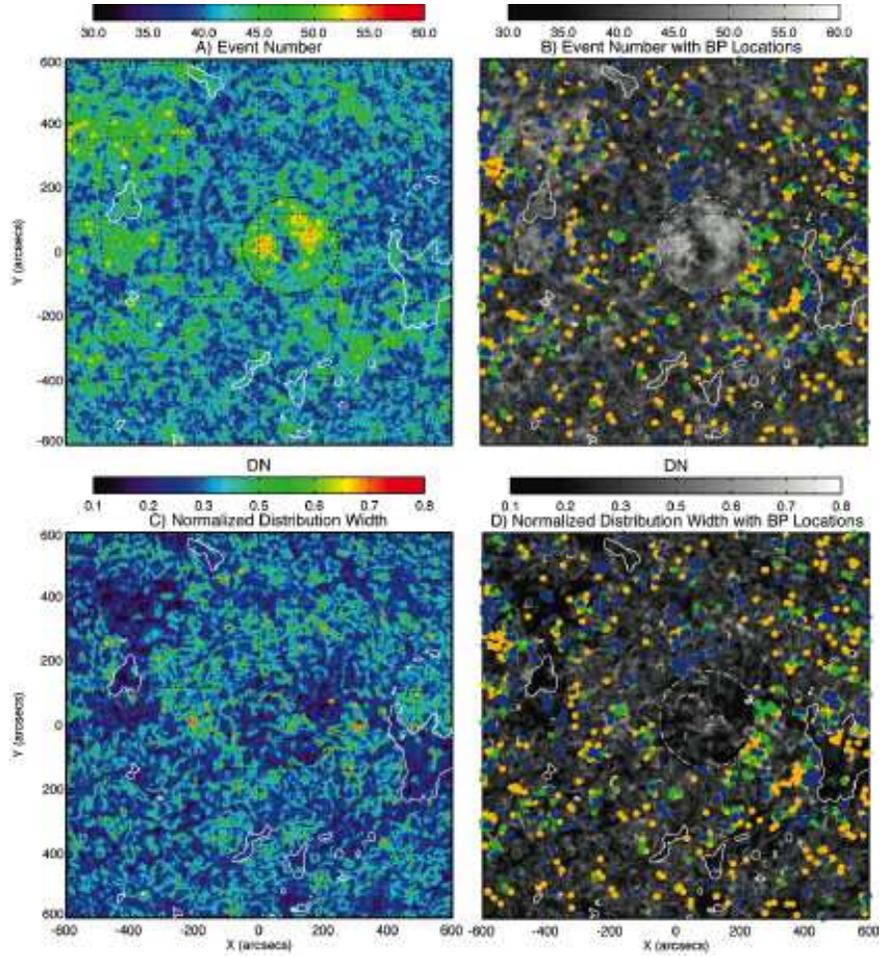}
\caption{Compare the event number map (top row) and normalized distribution width (bottom row) with the detected hot BP locations. The color of the dots indicates the EIT passband in which they were observed: blue (171\AA), green (195\AA) and yellow (284\AA). \label{res1}}
\end{figure}

Figure~\pref{res1} allows us to compare the event map (panel A) and normalized distribution width map (panel C) with the detected BP locations, panels B and D respectively. The color of the BPs shown on the right hand panels illustrate the EIT passband in which they were detected: blue from 171\AA; green from 195\AA;  and yellow from 284\AA. We notice immediately that the detected BP locations occur over regions where there are a large number of events (panel B) and the cavities occur in locations where there are small numbers of events. This is also consistent with the analysis of Paper~1 where it was noted that the brightest emission in the 304\AA{} passband occur in regions where the number of detected events are largest and the \brr{} is large (compare with panels A and C of Fig.~\pref{contextmarch2006}). We also see, from Panel D, that the BPs are offset from the distribution width (quiet Sun \-- supergranular interior) rings which is not surprising when the magnetic footpoints of BPs are at the vertices of the supergranulation pattern. The places where the normalized distribution width are largest, as we have seen in Fig.~\pref{bp1} (panel G), are additional markers of hot BPs. 

\begin{figure}
\plotone{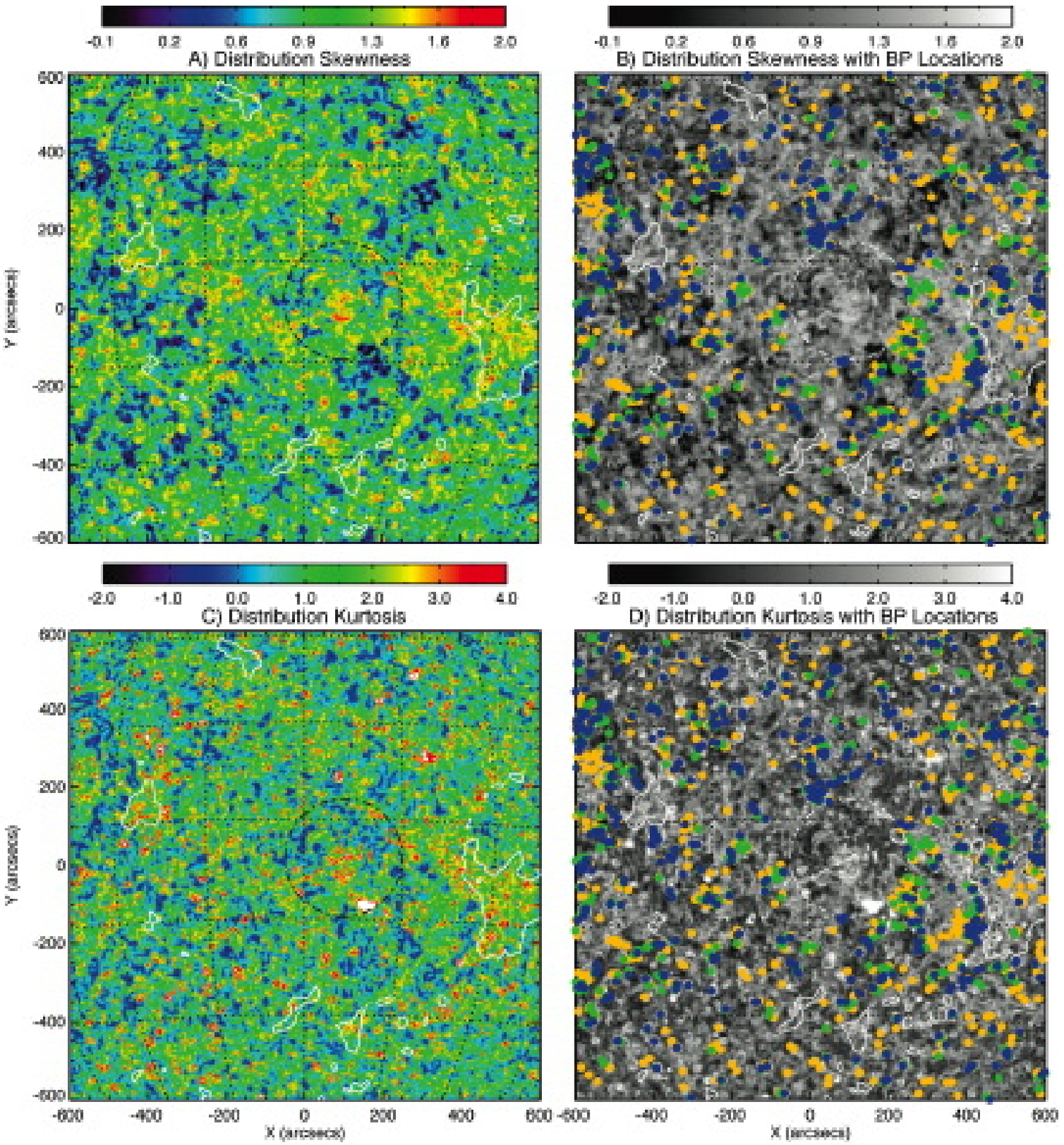}
\caption{Compare the distribution skewness (top row) and kurtosis (bottom row) with the detected hot BP locations. The color of the dots indicates the EIT passband in which they were observed: blue (171\AA), green (195\AA) and yellow (284\AA). \label{res2}}
\end{figure}

Figure~\pref{res2} compares the distribution skewness (panel A) and kurtosis (panel C) maps with the detected BP positions. We can clearly see that a (very) large number of the hot BPs do occur over skewness and kurtosis rings while there are locations where skewness an kurtosis rings exist, but hot BPs do not. This could either be caused by the lack of interaction (and reconnection) between the magnetic topology of the BP and the overlying coronal magnetic field \citep[discussed by][]{Harvey1994b} or that the 6-hour sampling of the EIT images causes us to miss some BPs that occur in the timeframe studied. The data analyzed lead us to believe that it is the former of these two possibilities; not all rings have hot BPs while all hot BPs have skewness and kurtosis rings associated with them. Why? We have seen above that BP cavities occur in locations of reduced (weaker than average quiet sun) coronal emission, but we must also consider the magnetic conditions at the root of the cavity and wether or not it will have the magnetic energy available to allow a hot BP to appear.

\begin{figure}
\plotone{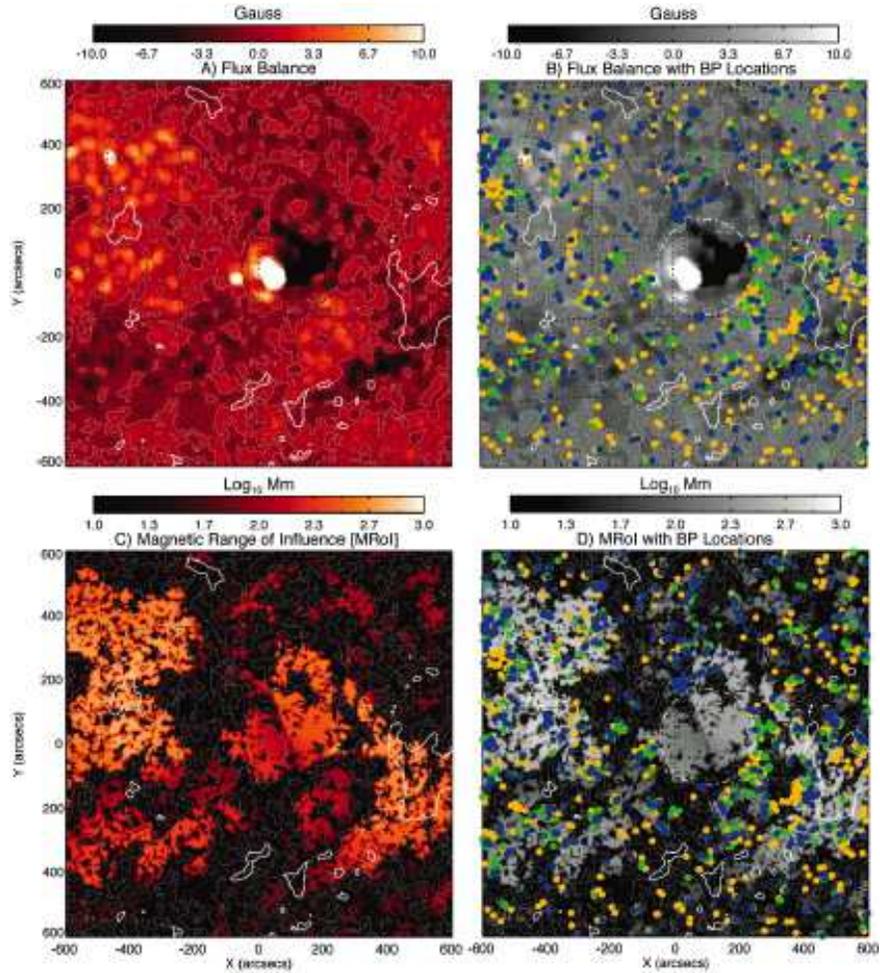}
\caption{Compare the supergranular averaged magnetic field strength (\brr; top row) and Magnetic Range of Influence (MRoI; bottom row) with the detected hot BP locations. The color of the dots indicates the EIT passband in which they were observed: blue (171\AA), green (195\AA) and yellow (284\AA). \label{res3}}
\end{figure}

Fig.~\pref{res3} illustrates the correspondence between the detected BP locations and the underlying magnetic field conditions: the \brr{} (panel A) and MRoI (panel C). We are instantly drawn to the fact that the large majority of BPs occur in places where both the absolute value of the \brr{} and MRoI are large, greater than $\sim$3G and $\sim$100Mm. Further, these comparative images indicate that very few hot BPs overlie magnetic neutral lines (regions of low \brr{} and MRoI). These are likely locations for BP cavities, regions where the magnetic field closes locally and has very little net imbalance. As we have suggested above, \citep[according to the results of][Paper~1]{McIntosh2006a, McIntosh2006b}, the corona above these regions is not very extended (low MRoI as a result of local magnetic closure) and does not receive as much mass or energy supply as the regions where the MRoI and \brr{} are large. The correspondence of BP cavities with regions of low MRoI and low imbalance lends further credibility to the deduction of \cite{Harvey1994b}; these regions of cool, weak corona cannot supply the additional mass and energy that is needed to generate and support the existence of hot BPs.

\section{Discussion}\label{discuss}
In the preceding sections we have observed that a simple analysis of EIT 304\AA{} passband image timeseries gives rise to several prevalent patterns in the derived diagnostics. This paper has concerned itself with the pattern that is closely associated with the appearance of EUV BPs; ``rings'' of enhanced distribution skewness and kurtosis. These rings span the vertices of one (or more) supergranules and occur as a consequence of abrupt emission changes in the region immediately around the BP. We have also noticed that the supergranular convective flow evolution governs the brightness fluctuations as the network vertices in these convective cells coalesce (or pull apart) as part of the natural flow evolution, or similarly as magnetic flux is being relentlessly driven to the boundary of the cells. In this case the appearance of BPs in the 304\AA{} passband is consistent with convection-forced magnetic reconnection at the interface between the chromosphere and transition region \citep[the likely formation region of the \ion{He}{2}, e.g.,][Paper~1]{JudgePietarila2004,PietarilaJudge2004}. We have noted that BPs have been linked with dynamic, magnetically forced phenomena \citep[][]{Moore1977, Habbal1981, Habbal1988, Porter1987, Habbal1991, Rovira1999, McQueen2000, Madjarska2002, Brosius2007} like those invoked by \citet{McIntosh2006b} to explain energy delivery to the quiet sun and coronal hole plasmas as a result of relentless magnetoconvection-forced reconnection \-- objects that look like (UV/EUV) spicules. Incidentally, the same forced reconnection should result in the leakage of {\it p-modes} into the closed magnetic system \citep[e.g.,][]{DePontieu2004a, DePontieu2004b, McIntosh2006, Hansteen2006, Jefferies2006, DePontieu2007} and explain the observation of propagating 3 and 5-minute waves in BPs \citep[e.g.,][]{Urra2004a}.

We have discriminated between these cool (visible in the 304\AA{} passband) and hot EIT BPs \citep[visible in the 171, 195 or 284\AA{} passbands, detected using the method of][]{McIntoshGurman2005}. We have noticed that not all cool BPs have an associated hot BP, as was also noted by \citet{Harvey1994a} and others. That is, while all cool BPs have skewness and kurtosis rings associated with them, not all rings have hot BPs overlying them. Additionally, we have noted the presence of regions in the hot corona where very few hot BPs occur and have dubbed them ``BP cavities''. These BP cavities are co-spatial with locations in the corona where the quiet Sun emission is reduced, but to the level found in a coronal hole. This last point indicates that the global magnetic field topology in the BP cavity is closed (low \brr{} and low MRoI) and that the energy supply to the coronal plasma surrounding the 304 BP is generally reduced \citep[e.g.,][]{McIntosh2006b}. It is not unreasonable, therefore, to expect that this reduction in global energy input is an important factor in explaining why some cool BPs do not become hot BPs. To this end, we propose that (both cool and hot) BPs are subject to a two-stage heating process, one forced by the constantly evolving magnetic flux elements rooted in the high-$\beta$ convective flow as we have discussed above and a second stage, that only appears to be significant when the BP is immersed in an energetic magnetic environment (high \brr{}, moderate to high MRoI). In the latter case, following the presented analysis and that of Paper~1, stage one will be very vigorous, continuously dumping mass and energy into the BP. As the stage one loading progresses, the BP will expand vertically and begin to interact with the energetic corona that surrounds it (stage two), amplifying its brightness in a manner described by \citet{Longcope2001, Brown2001, Brown2002}. 

Unfortunately, it remains to be seen if the enhanced distribution widths at the BP apex (e.g., Fig.~\pref{bp1}G) are the direct signature of the stage-two connection between the overlying corona and the expanding BP; this is an issue that we also intend to follow-up in future work. Similarly, we have seen that the evolution of BPs is a very complex magnetic interaction with a basis that hinges on the supergranular evolution and magnetic flux aggregation on that spatial scale \citep[e.g.,][]{Crouch2007}. We can say then that the dynamic evolution of the individual supergranular cells is itself the controlling factor in establishing BPs lifetimes. 

Finally, in Sect.~\pref{single} we demonstrated the complex evolution of a typical BP from birth to its eventual death as an monopolar object. Using this BP as an example we can speculate on what might happen to a similar BP that it is now anchored in a coronal hole (a open magnetic region with high MRoI and non-negligable magnetic imbalance, non-zero \brr). We allude to the possible conversion of coronal hole BPs into polar plumes \citep[e.g.,][]{DeForest1997,DeForest1998,Wang1998}. We propose that this is the natural consequence of the BP's evolution from a dense, bright dipolar to a bright, effectively monopolar, straw-like coronal structure \citep[typical magnetic network elements in coronal holes are not bright in the EUV while plumes are, e.g.,][]{McIntosh2006b}. The evolution from BP to plume occurs because one BP footpoint is constantly being eroded by the imposed net imbalance of the coronal hole itself, launching compressive waves and dense plasma into the fast solar wind. It is precisely the evolution of the single BP and its impact on the energetics of the solar wind and corona that are of interest for {\em Hinode} \citep[][]{Ichimoto2005} and subsequent flights of the EUNIS Sounding Rocket \citep[][]{Thomas2001,Brosius2007}. In particular, the instrument suite of {\em Hinode} will be able to monitor the evolution of the photospheric magnetic field simultaneously with the response of the chromospheric, transition region and coronal plasmas at unprecidented spatial and temporal resolution. These observations will allow a significant advance in the analysis and interpretation of the modest resolution data presented here. They will allow us to investigate the impact of forced magnetic reconnection and the subsequently mediated {\em p-mode} leakage on the BP structure and the role that the surrounding magnetic environment plays in the appearance of the BP in the corona as a microcosm of the global magnetic corona.

\section{Conclusion}
We have used long-duration observations in the 304\AA{} passband of EIT, with supporting data from MDI and the EIT Bright Point database (McIntosh \& Davey 2007 \-- in preparation), to discuss the physical processes and conditions that support the appearance and evolution of hot (EUV, X-Ray) Bright Points (BP). We believe that these data are consistent with the picture that mass and energy input into the BPs comes from the magnetic reconnection forced upon the BP by the high-$\beta$ flow on the magnetic flux elements at the vertices of supergranular boundaries and the ongoing destruction of advecting magnetic flux around these vertices, consistent with the recent analysis of \citep{Brosius2007}. Further, we have seen that production of a hot BP relies on the magnetic topology of the overlying corona that is imposed by the ``footprint'' of the photospheric magnetic field (the \brr{} and MRoI) on the plasma. The resulting effect is that hot BP production is a two-stage process. The first (and most prominent) stage is that involving the injection of mass and energy through forced magnetoconvection-forced reconnection \citep[also observed by][]{Moore1977, Rovira1999, Madjarska2002}, while the second occurs when there is enough energy supply to the BP that it will develop an expanded loop system that begins to interact with the overlying corona through separator reconnection \citep[e.g.,][]{Longcope2001, Brown2001, Brown2002} further increasing the plasma temperature at which it becomes visible. 

\acknowledgements 
The author would like to thank Bart DePontieu, Joan Burkepile, Alisdair Davey, Joe Gurman, Phil Judge and Meredith Wills-Davey for very helpful discussions about this manuscript and to the folks at HAO for their hospitality. This material is based upon work carried as a visitor to NCAR and at the Southwest Research Institute. This work was supported by the National Aeronautics and Space Administration under grants to the author issued by \soho{} (NNG05GQ70G), the Sun-Earth Connection Guest Investigator Program (NNG05GM75G), the Solar and Heliospheric Physics Program (NNG06GC89G) and the National Science Foundation Solar Terrestrial Physics Program (ATM-0541567). The National Center for Atmospheric Research is sponsored by the National Science Foundation.

\clearpage

\begin{thebibliography}{72}

\bibitem[{{Abramowitz} \& {Stegun}(1972)}]{Abramowitz1972}
Abramowitz, M. \& Stegun, I. A. (Eds.)., 1972, Handbook of Mathematical Functions with Formulas, Graphs, and Mathematical Tables, 9th printing. New York: Dover, 920

\bibitem[{{Andretta} {et~al.}(2000)}]{Andretta2000}
Andretta, V., et~al., 2000, \apj, 535, 438

\bibitem[{{Athay}(2002)}]{Athay2002}
{Athay}, R.~G., 2002, \solphys, 197, 31

\bibitem[{{Aschwanden}, {Winebarger} \& {Tsiklauri}(2007)}]{Aschwanden2007}
{Aschwanden}, M.~J., {Winebarger}, A. \& {Tsiklauri}, D., ``The Coronal Heating Paradox'', Submitted \apj{} Lett. (October 2006)

\bibitem[{{Beckers}(1968)}]{Beckers1968}
{Beckers}, J.~M., 1968, \solphys, 3, 367

\bibitem[{{Brosius}, {Davila} \& {Thomas}(1998)}]{Brosius1998}
{Brosius}, J.~W., {Davila}, J.~M. \& {Thomas}, R.~J., 1998, \apjs, 119, 255

\bibitem[{{Brosius}, {Rabin} \& {Thomas}(2007)}]{Brosius2007}
{Brosius}, J.~W., {Rabin}, D.~M. \& {Thomas}, R.~J., 2007, \apj, 656, L41

\bibitem[{{Brown}(2001)}]{Brown2001}
{Brown}, D.~S., et~al. 2001, \solphys, 201, 305

\bibitem[{{Brown}(2002)}]{Brown2002}
{Brown}, D.~S., et~al. 2002, Ad.~Sp.~Res., 29, 1093

\bibitem[{{Cattaneo} {et~al.}(2003)}]{Cattaneo2003}
{Cattaneo}, F., {Emonet}, T. \& {Weiss}, N. 2003, \apj, 588, 1183

\bibitem[{{Crouch}, {Charbonneau} \& {Thibault}(2007)}]{Crouch2007}
Crouch, A.D., Charbonneau, P. \& Thibault, K., in press \apj (Feb. 2007)

\bibitem[{{DeForest} {et~al.}(1997)}]{DeForest1997}
{DeForest}, C.~E. et~al. 1997, \solphys, 175, 393

\bibitem[{{DeForest} \& {Gurman}(1998)}]{DeForest1998}
{DeForest}, C.~E. \& {Gurman}, J.~B. 1998, \apjl, 501, 217

\bibitem[{{Delaboudini\`{e}re} {et~al.}(1995)}]{Boudine1995}
{Delaboudini\`{e}re}, J.-P. {et~al.} 1995, \solphys, 162, 291

\bibitem[{{De Pontieu} {et~al.}(2004a){De Pontieu}, {Erdelyi}, \& {James}}]{DePontieu2004a}
{De Pontieu}, B., {Erdelyi}, R., {James}, S.P., 2004a, Nature, 430, 536

\bibitem[{{De Pontieu} {et~al.}(2004b){De Pontieu}, {Erdelyi}, \& {De Moortel}}]{DePontieu2004b}
{De Pontieu}, B., {Erdelyi}, R., {De Moortel}, I., 2004b, \apjl, 624, 61

\bibitem[{{De Pontieu} {et~al.}(2007a)}]{DePontieu2007}
{De Pontieu}, B., et~al., 2007a, \apj, 655, 624

\bibitem[{{De Pontieu} {et~al.}(2007b)}]{DePontieu2007b}
{De Pontieu}, B.,  et~al., 2007b, in {\it Coimbra Solar Physics Meeting on the Physics of Chromospheric Plasmas}, ASP Conference Series, 354, ed. P. Heinzel, I. Dorotovic, R.J. Rutten, in press (also at astro-ph/0702081)

\bibitem[{{Freeland} \& {Handy}(1998)}]{Freeland1998} 
Freeland, S.~L. and Handy, B.~N. 1998, \solphys, 182, 497

\bibitem[{{Finsterle} {et~al.}(2004a)}]{Finsterle2004a}
{Finsterle}, W., {Jefferies}, S.~M., {Cacciani}, A., et~al. 2004a, \solphys, 220, 317

\bibitem[{{Fleck} {et~al.}(1995){Fleck}, {Domingo}, \& {Poland}}]{Fleck1995}
{Fleck}, B., {Domingo}, V., {Poland}, A.~I. 1995, {\em The SOHO mission},
(Dordrecht: Kluwer)

\bibitem[{{Golub} {et~al.}(1974)}]{Golub1974} 
Golub, L., et~al. 1974, \apjl, 189, L93

\bibitem[{{Habbal} \& {Withbroe}(1981)}]{Habbal1981}
Habbal, S.~R. \& Withbroe, G.~L., 1981, \solphys, 77, 97

\bibitem[{{Habbal} \& {Harvey}(1988)}]{Habbal1988}
Habbal, S.~R. \& Harvey, K.~L., 1988, \apj, 326, 988

\bibitem[{{Habbal} \& {Grace}(1991)}]{Habbal1991}
Habbal, S.~R. \& Grace, E., 1991, \apj, 382, 667

\bibitem[{{Hale} {et~al.}(1919)}]{Hale1919}
Hale, G.~E., et~al. 1919, \apj, 49, 153

\bibitem[{{Hansteen} {et~al.}(2006)}]{Hansteen2006}
{Hansteen}, V.~H., et~al. 2006, \apj, 647, L73

\bibitem[{{Harvey}(1993)}]{Harvey1993} 
Harvey, K.~L., et al. 1993, Ad. Sp. Res., 13, 27

\bibitem[{{Harvey}(1994a)}]{Harvey1994a} 
Harvey, K.~L., 1994, In ``Infrared Solar Physics'', IAU Symposium 154, Rabin, D.M. et al. (Eds.), 71

\bibitem[{{Harvey}(1994b)}]{Harvey1994b} 
Harvey, K.~L., 1997, In ``Solar Active Region Evolution: Comparing Models with Observations''. ASP Conference Series; Vol. 68, Balasubramaniam, K.S. \& SImon, G.W. (Eds.), 377

\bibitem[{{Harvey}(1997)}]{Harvey1997} 
Harvey, K.~L., 1997, In ``Magnetic Reconnection in the Solar Atmosphere''. ASP Conference Series; Vol. 111, Bentley R.~D. \& Mariska J.~T. (Eds.), 9

\bibitem[{{Ichimoto} {et~al.}(2005)}]{Ichimoto2005}
Ichimoto, K. et~al., 2005, J. Kor. Ast. Soc., 38, 307

\bibitem[{{Jefferies} {et~al.}(2006)}]{Jefferies2006}
Jefferies, S.~M., et~al. 2006, \apjl, 648, L151

\bibitem[{{Jordan}(1975)}]{Jordan1975}
Jordan, C., 1975, MNRAS, 170, 429

\bibitem[{{Judge} \& {Pietarila}(2004)}]{JudgePietarila2004}
Judge, P.~G., Pietarila, A., 2004, 606, 1258

\bibitem[{{Klimchuk}(2006)}]{Klimchuk2006}
Klimchuk, J.~A. 2006, \solphys, 234, 41

\bibitem[{{Lin} {et~al.}(2003)}]{Lin2003}
Lin, G. et~al., 2003, Cytometry, Part A, 56A, 23

\bibitem[{{Lin} {et~al.}(2005)}]{Lin2005}
Lin, G. et~al., 2005, Cytometry, Part A, 63A, 20

\bibitem[{{Longcope} {et~al.}(2001)}]{Longcope2001}
Longcope, D.~W., et~al., 2001, \apj, 553, 429

\bibitem[{{L\'{o}pez Fuentes}, {Klimchuk} \& {D\'{e}moulin}(2006)}]{Lopes2006}
{L\'{o}pez Fuentes}, M.~C., {Klimchuk}, J.~A. \& {D\'{e}moulin}, P., \apj, 639, 459

\bibitem[{{Madjarska} \& {Doyle}(2002)}]{Madjarska2002}
Madjarska, M.~S. \& Doyle, J.~G., 2002, In Proc. SOHO 11, ``From Solar Min to Max: Half a Solar Cycle with SOHO'', 11-15 March 2002, Davos. A. Wilson (Ed.), ESA SP-508, 311

\bibitem[{{Mauas} {et~al.}(2005)}]{Mauas2005}
Mauas, P.~J.~D., 2005, \apj, 619, 604

\bibitem[{{Mazzotta} {et~al.}(1998)}]{Mazzotta1998}
Mazzotta, P., Mazzitelli, G., Colafrancesco, S., \& Vittorio, N. 1998, \aaps, 133, 403

\bibitem[{{McIntosh} {et~al.}(2004){McIntosh}, Fleck, \& Tarbell}]{McIntosh2004}
McIntosh, S.~W., Fleck, B., \& Tarbell, T.~D. 2004, \apjl, 609, L95

\bibitem[{{McIntosh} \& {Gurman}(2005)}]{McIntoshGurman2005}
McIntosh, S.~W. \& Gurman, J.~B., 2005, \solphys, 228, 285

\bibitem[{{McIntosh} \& {Jefferies}(2006)}]{McIntosh2006}
{McIntosh}, S.~W. \& {Jefferies}, S.~M. 2006, \apj, 648, L151 

\bibitem[{{McIntosh} {et~al.}(2006)}]{McIntosh2006a}
McIntosh, S.~W., Davey, A.R., \& Hassler D.~M. 2006, \apj, 644, L87

\bibitem[{{McIntosh} {et~al.}(2007a)}]{McIntosh2006b}
McIntosh, S.~W., et~al., 2007a, \apj, 654, 650

\bibitem[{{McIntosh} {et~al.}(2007b)}]{McIntosh+others2007a}
McIntosh, S.~W. et~al., 2007b, \apj, (in press)

\bibitem[{{McIntosh}(2007)}]{McIntosh2007}
McIntosh, S.~W., 2007, \apj, 657, L125 (Paper~1)

\bibitem[{{McIntosh}, {Judge} \& {Gurman}(2007)}]{McIntosh+others2007b}
McIntosh, S.~W., Judge, P.~G. \& Gurman, J.~B. 2007, ``On the Magnetoconvective Occurrence of Spicules: The Fundamental Building Block of Solar Energy Release'', In prep \apj.

\bibitem[{{MacQueen} {et~al.}(2000)}]{McQueen2000}
MacQueen, R.M., et~al., 2000, \solphys, 191, 2000

\bibitem[{{Moore} {et~al.}(1977)}]{Moore1977}
Moore, R.~L., et~al., 1977, \apj{}, 218, 286

\bibitem[{{Parker}(1964)}]{Parker1964}
{Parker}, E.~N., 1964, \apj, 140, 1170

\bibitem[{{Parker}(1988)}]{Parker1988}
{Parker}, E.~N., 1988, \apj, 330, 474

\bibitem[{{Parker}(1994)}]{Parker1994}
Parker, E.~N., 1994 ÒSpontaneous Current Sheets in Magnetic Fields: with applications to stellar x-raysÓ, Oxford University Press

\bibitem[{{Pietarila} \& {Judge}(2004)}]{PietarilaJudge2004}
Pietarila, A., Judge, P.~G. 2004, 606, 1239

\bibitem[{{Porter} {et~al.}(1987)}]{Porter1987}
Porter, J.~G., Moore, R.~L., Reichmann, E.~J. \& Harvey, K.L., 1987, \apj, 323, 380

\bibitem[{{Press} {et~al.}(1992)}]{Press1992}
Press, W.~H. et~al. 1992, Numerical Recipes in FORTRAN: The Art of Scientific Computing, 2nd ed. Cambridge, England: Cambridge University Press

\bibitem[{Priest {et~al.}(2002){Priest}, {Heyvaerts}, \& {Title}}]{Priest2002}
{Priest}, E.~R., {Heyvaerts}, J.~F. \& {Title}, A.~M. 2002, \apj, 576, 533

\bibitem[{{Roberts}(1945)}]{Roberts1945}
Roberts, W.~O. 1945, \apj, 101, 136

\bibitem[{{Rovira} {et~al.}(1999)}]{Rovira1999}
Rovira, M., {et~al.} 1999, \apj, 510, 474

\bibitem[{Scherrer {et~al.}(1995)}]{Scherrer1995}
{Scherrer}, P.~H., {et~al.} 1995, \solphys, 162, 129

\bibitem[{{Schrijver} {et~al.}(1997)}]{Schrijver1997}
Schrijver, C.~J., et~al. 1997, \apj, 487, 424

\bibitem[{{Secchi}(1877)}]{Secchi1877}
Secchi, P.~S. 1877, Le Soliel, Paris, Gaultier-Villars

\bibitem[{{Thomas} \& {Davila}(2001)}]{Thomas2001}
Thomas, R.~J. \& Davila, J.~M., 2001, Proc. SPIE, 4498, 161.

\bibitem[{{Ugarte-Urra}(2004)}]{Urra2004}
Ugarte-Urra, I., 2004, PhD. Thesis, ``Brightness and Magnetic Evolution of Solar Coronal Bright Points'', Queen's University of Belfast, UK

\bibitem[{{Ugarte-Urra} {et~al.}(2004a)}]{Urra2004a}
Ugarte-Urra, I., et~al., 2004a, \aap, 418, 313

\bibitem[{{Ugarte-Urra} {et~al.}(2004b)}]{Urra2004b}
Ugarte-Urra, I., et~al., 2004b, \aap, 425, 1083

\bibitem[{{Ugarte-Urra} \& {Doyle}(2004)}]{UrraDoyle2004}
Ugarte-Urra, I., Doyle, J.~G., 2004, In Proc. SOHO 15 - ``Coronal Heating''. 6-9 September 2004, R.W. Walsh, J. Ireland, D. Danesy, B. Fleck. (Eds.), ESA SP-575, 535

\bibitem[{{Ugarte-Urra}, {Doyle} \& {Del Zanna}(2005)}]{Urra2005}
Ugarte-Urra, I., Doyle, J.~G., Del Zanna, G., 2005, \aap, 435, 1169

\bibitem[{{Wang}(1998)}]{Wang1998}
Wang, Y.-M., 1998, \apjl, 501, 145

\bibitem[{{Zirin}(1975)}]{Zirin1975}
Zirin, H., 1975, \apj, 199, L63

\end{thebibliography}
\end{document}